\documentclass[11pt]{article}
\usepackage{authblk}
\usepackage{graphicx}
\usepackage{epstopdf}
\usepackage{latexsym}
\usepackage{multirow}
\usepackage{tabls}
\usepackage{epstopdf}
\usepackage{graphicx}
\usepackage{amsmath, amsthm, amssymb}
\title{Complex Representation of Potentials and Fields for the Nonlinear Magnetic Insert of the Integrable Optics Test Accelerator}
\author{Chad Mitchell}
\affil{{Lawrence Berkeley National Laboratory, Berkeley, California, 94720, USA}}
\begin{document}
\maketitle
\begin{abstract}
An alternative representation for the vector potential of the nonlinear magnetic insert for the Integrable Optics Test Accelerator (IOTA), first described in Sec. V.A. of \cite{Danilov}, is determined from first principles using standard complex variable methods.  In particular, it is shown that the coupled system consisting of the 2D Laplace equation and the Bertrand-Darboux equation is equivalent to a single ordinary differential equation in the complex plane, and a simple solution is constructed.  The results are consistent with \cite{Danilov}, and this concise representation provides computational advantages for particle tracking through the nonlinear insert by avoiding numerical errors caused by small denominators that appear when evaluating transverse derivatives of the vector potential near the midplane.  A similar representation is provided for the spatial dependence of the two invariants of motion.
\end{abstract}

\section{Introduction}
The Integrable Optics Test Accelerator (IOTA), currently under construction at Fermilab, is a novel storage ring designed in part to investigate the dynamics of beams with large transverse tune spread in the presence of a strongly nonlinear integrable lattice \cite{Valishev}.  The lattice design, first described in \cite{Danilov}, makes use of a 2 m-long nonlinear magnetic insert with an $s$-dependent transverse magnetic field that is shaped to generate bounded, regular (integrable) motion in the transverse plane for on-momentum particles.  One primary goal of IOTA is to achieve nonlinear tune shifts exceeding 0.25 without significant degradation of the dynamic aperture \cite{Valishev}, while the strong dependence of particle tune on amplitude produces decoherence of transverse oscillations that may help to prevent the development of coherent space charge instabilities \cite{Webb,Webb2}.

The vector potential within the nonlinear insert, as described in \cite{Danilov}, is expressed using a variable transformation in the transverse plane of the following form:
\begin{equation}
\xi=\frac{\sqrt{(x+f)^2+y^2}+\sqrt{(x-f)^2+y^2}}{2f},\quad\quad \eta=\frac{\sqrt{(x+f)^2+y^2}-\sqrt{(x-f)^2+y^2}}{2f}, \label{transformation}
\end{equation}
where $f>0$ is an arbitrary constant.  The Jacobian determinant of this transformation is given by: 
\begin{equation}
J=-\frac{y}{f\sqrt{(x-f)^2+y^2}\sqrt{(x+f)^2+y^2}},
\end{equation}
which vanishes when $y=0$.  It follows that the transformation of first-order derivatives from the basis $(\partial_x,\partial_y)$ to $(\partial_{\xi},\partial_{\eta})$ under (\ref{transformation}) is not invertible when $y=0$.  As a consequence, expressions for the first-order derivatives of the vector potential, which appear in the equations of motion, have denominators that vanish in the midplane.  Successful numerical tracking through the nonlinear insert therefore requires the use of Taylor approximation near $y=0$ to avoid introducing large numerical errors \cite{CHall}.  The resulting expressions are lengthy, subject to human error during implementation, and may result in a loss of numerical accuracy over $\sim$100's of turns.

In this document, an alternative, complex-variables approach to the nonlinear insert is described that avoids the transformation (\ref{transformation}) and the resulting growth of numerical errors.  Much of the theoretical development in Sec. V.A. of \cite{Danilov} is revisited from a slightly different perspective.

\section{Integrability and the Bertrand-Darboux Equation}
Following \cite{Danilov}, we search for a magnetic vector potential such that the transverse Hamiltonian for an on-energy particle within this vector potential takes the form:
\begin{equation}
H_{\perp}(X,P_x,Y,P_y;s)=\frac{1}{2}(P_x^2+P_y^2)-\frac{\tau c^2}{\beta(s)}U\left(\frac{X}{c\sqrt{\beta(s)}},\frac{Y}{c\sqrt{\beta(s)}}\right), \label{Hamiltonian}
\end{equation}
where $P_x=p_x/p^0$ and $P_y=p_y/p^0$ are momenta normalized by the total design momentum $p^0$, $\beta=\beta_x=\beta_y$ is the betatron amplitude across the drift space that will contain the magnet, and 
$U$ is a 2D harmonic solution of the Bertrand-Darboux equation:
\begin{equation}
(\partial_x^2+\partial_y^2)U=0,\quad\quad xy(\partial_{x}^2-\partial_{y}^2)U+(y^2-x^2+1)\partial_{x}\partial_yU+3y\partial_xU-3x\partial_yU=0. \label{Uproperties}
\end{equation}
We use $s$ to denote the longitudinal coordinate throughout, in order to avoid confusion with a complex variable $z$ to be defined in Section 4.  The parameters $\tau$ and $c$ have been chosen according to standard usage, so that $\tau$ is a dimensionless parameter characterizing the strength of the nonlinear insert, and $c\neq 0$ [m$^{1/2}$] characterizes the length scale of the potential in the transverse plane.

To motivate conditions (\ref{Hamiltonian}-\ref{Uproperties}), we perform an $s$-dependent Courant-Snyder transformation of the phase space variables similar to that described in Eq. (2) of \cite{Danilov}, to new variables given by:
\begin{equation}
\begin{pmatrix}
X_N \\ P_{xN}
\end{pmatrix}=
\begin{pmatrix}
1/c\sqrt{\beta} & 0 \\
\alpha/c\sqrt{\beta} & \sqrt{\beta}/c  \\
\end{pmatrix}
\begin{pmatrix}
X \\ P_x 
\end{pmatrix},\quad\quad
\begin{pmatrix}
Y_N \\ P_{yN}
\end{pmatrix}=
\begin{pmatrix}
1/c\sqrt{\beta} & 0 \\
\alpha/c\sqrt{\beta} & \sqrt{\beta}/c  \\
\end{pmatrix}
\begin{pmatrix}
Y \\ P_y 
\end{pmatrix}.\label{normalizing}
\end{equation}
Here $\alpha(s)=-\beta'(s)/2$ denotes the usual Twiss function, and the factors of $c$ are included so that the normalized variables $(X_N,P_{xN},Y_N,P_{yN})$ are dimensionless.
Taking the betatron phase advance $\psi$ (defined by $\psi'=1/\beta$) as the independent variable, the Hamiltonian corresponding to (\ref{Hamiltonian}) in these normalized variables takes the form:
\begin{equation}
H_N(X_N,P_{xN},Y_N,P_{yN})=\frac{1}{2}(P_{xN}^2+P_{yN}^2+X_N^2+Y_N^2)-\tau U\left(X_N,Y_N\right). \label{Htransform}
\end{equation}
Since (\ref{Htransform}) no longer contains the independent variable, $H_N$ is an invariant of motion.  Consider the following ansatz for a second invariant of motion:
\begin{equation}
I_N(X_N,P_{xN},Y_N,P_{yN})=(X_NP_{yN}-Y_NP_{xN})^2+P_{xN}^2+X_N^2-\tau W(X_N,Y_N), \label{Itransform}
\end{equation}  
where $W$ is some smooth function of coordinates only.
Taking the Poisson bracket of (\ref{Htransform}) and (\ref{Itransform}) and collecting terms by degree in the momenta gives:
\begin{equation}
\left\{H_N,I_N\right\}=\tau P_{xN}\left(W_x-2Y_N^2U_x-2U_x+2X_NY_NU_y\right)+\tau P_{yN}\left(W_y+2X_NY_NU_x-2X_N^2U_y\right), \label{PB}
\end{equation}
where a subscript on $U$ or $W$ denotes that a partial derivative is taken along the indicated normalized coordinate.  Clearly (\ref{PB}) vanishes when $\tau=0$.
When $\tau\neq 0$, (\ref{PB}) vanishes if and only if $U$ and $W$ are connected through the pair of partial differential equations:
\begin{align}
W_x=2(Y_N^2+1)U_x-2X_NY_NU_y,\quad\quad W_y=-2X_NY_N U_x+2X_N^2 U_y. \label{Weq}
\end{align}
Taking the $y$-derivative of the leftmost equation in (\ref{Weq}) and the $x$-derivative of the rightmost equation in (\ref{Weq}) and subtracting the two allows us to eliminate the dependence on $W$, and the result is the Bertrand-Darboux equation \cite{Darboux}-\cite{Darboux2} for an integrable potential $U$ (the rightmost equation in (\ref{Uproperties})).  Given a function $U$ satisfying this equation, (\ref{Weq}) can be solved to determine the corresponding function $W$, and therefore a second invariant of motion.  Since (\ref{Htransform}) and (\ref{Itransform}) are also functionally independent, it follows that the portion of the ring represented by (\ref{Hamiltonian}) is described by a completely integrable Hamiltonian system.

Suppose that the transfer map $\mathcal{R}$ for the remainder of the ring (from the exit of the nonlinear insert to its entrance) is described in normalized coordinates (\ref{normalizing}) by a matrix $\pm I$, where $I$ is the 4$\times$4 identity.  We will see (\ref{Uresult}, \ref{WResult2}) that $U$ and $W$ are each symmetric under inversion of the dynamical variables.  It will then follow from the inversion symmetry of (\ref{Htransform}-\ref{Itransform})  that $H_N$ and $I_N$ are each invariant under the map $\mathcal{R}$, and therefore under the one-turn map for the ring.  Thus, the complete ring is described by an integrable symplectic map.

To ensure integrability of motion in such a ring, it remains to design a magnetic vector potential such that the Hamiltonian $H_{\perp}$ within the nonlinear insert satisfies (\ref{Hamiltonian}-\ref{Uproperties}).  Background for this problem is considered in Section 3.  In Section 4, the system (\ref{Uproperties})  is solved to determine the function $U$ appearing in the Hamiltonian, and the magnetic potentials and magnetic fields are described.  In Section 5, the system (\ref{Weq}) is solved to determine the function $W$ appearing in the second invariant of motion.  Section 6 summarizes the primary results for reference.

\section{Hamiltonian and Comments on the Vector Potential}
We begin with the single-particle Hamiltonian $K$ for a magnetostatic element with the longitudinal coordinate $s$ as the independent variable, in the form \cite{Dragt}:
\begin{equation}
K(x,p_x,y,p_y,t,p_t;s)=-\sqrt{\frac{p_t^2}{c^2}-m^2c^2-(\vec{p}_{\perp}-q\vec{A}_{\perp})^2}-qA_s.
\end{equation}
In this section, note that $c$ denotes the speed of light, and $\beta_0$, $\gamma_0$ will denote the relativistic factors at the design energy.
It is convenient to work in dimensionless variables defined by a scale length $l$, a scale momentum $p^0$, and a scale frequency $\omega=c/l$, where $p^0=\gamma_0\beta_0mc$ is the design momentum.  Thus we use the variables $\bar{x}=x/l$, $\bar{p}_x=p_x/p^0$, $\bar{y}=y/l$, $\bar{p}_y=p_y/p^0$, $\bar{t}=ct/l$, and $\bar{p}_t=p_t/(cp^0)$, and the Hamiltonian in these variables becomes \cite{Ryne}:
\begin{equation}
\bar{K}(\bar{x},\bar{p}_x,\bar{y},\bar{p}_y,\bar{t},\bar{p}_t;s)=-\frac{1}{l}\sqrt{\bar{p}_t^2-\left(\frac{mc}{p^0}\right)^2-\left(\vec{\bar{p}}_{\perp}-\frac{q}{p^0}\vec{A}_{\perp}\right)^2}-\frac{q}{lp^0}A_s. \label{Kdim}
\end{equation}
For convenience, we set $l=1$ m and we define a dimensionless vector potential $\vec{\mathcal{A}}=\vec{A}/B\rho=(q/p^0)\vec{A}$.  Since $\bar{K}$ is independent of $\bar{t}$, it follows that $\bar{p}_t$ is a constant of motion, with design value $\bar{p}_t^r=-\gamma_0mc^2/(cp^0)=-1/\beta_0$.  If $\vec{A}_{\perp}$ and $\nabla_{\perp}A_s$ both vanish along the axis, it follows from (\ref{Kdim}) that a valid trajectory through the magnet is given by the on-axis orbit $\bar{x}^r=\bar{y}^r=\bar{p}_x^r=\bar{p}_y^r=0$.  It also follows that this trajectory satisfies $d\bar{t}^r/ds=1/\beta_0$.  Defining the deviation variables $X=\bar{x}-\bar{x}^r$, $P_x=\bar{p}_x-\bar{p}_x^r$, $Y=\bar{y}-\bar{y}^r$, $P_y=\bar{p}_y-\bar{p}_y^r$, $T=\bar{t}-\bar{t}^r$, and $P_t=\bar{p}_t-\bar{p}_t^r$, motion in these deviation variables is described by the Hamiltonian \cite{Ryne}:
\begin{equation}
H(X,P_x,Y,P_y,T,P_t;s)=-\sqrt{\left(P_t-\frac{1}{\beta_0}\right)^2-\frac{1}{(\beta_0\gamma_0)^2}-\left(\vec{P}_{\perp}-\vec{\mathcal{A}}_{\perp}\right)^2}-\mathcal{A}_s-\frac{1}{\beta_0}\left(P_t-\frac{1}{\beta_0}\right). \label{Kdev}
\end{equation}
The constant term (rightmost) can be dropped without affecting the equations of motion, and simplifying slightly we have:
\begin{equation}
H(X,P_x,Y,P_y,T,P_t;s)=-\sqrt{1-\frac{2P_t}{\beta_0}+P_t^2-\left(\vec{P}_{\perp}-\vec{\mathcal{A}}_{\perp}\right)^2}-\mathcal{A}_s-\frac{1}{\beta_0}P_t. \label{Kdev2}
\end{equation}
So far, we have made no approximations.  To proceed further, note that \cite{Danilov} is concerned only with the transverse motion of on-energy particles.  Since $P_t$ is a constant of motion, we set $P_t=0$ to obtain the transverse Hamiltonian for on-energy particles:
\begin{equation}
H_{\perp}(X,P_x,Y,P_y;s)=-\sqrt{1-\left(\vec{P}_{\perp}-\vec{\mathcal{A}}_{\perp}\right)^2}-\mathcal{A}_s. \label{Kdev3}
\end{equation}
The term in parentheses is the just the transverse mechanical momentum (normalized by the design momentum $p^0$).  Under the paraxial approximation that the transverse mechanical momentum is much smaller than the longitudinal momentum, we may expand the square root and neglect any residual constant terms to obtain the approximate transverse Hamiltonian for on-energy particles:
\begin{equation}
H_{\perp}(X,P_x,Y,P_y;s)\approx \frac{1}{2}\left(\vec{P}_{\perp}-\vec{\mathcal{A}}_{\perp}\right)^2-\mathcal{A}_s. \label{KdevAppx}
\end{equation}
Comparing (\ref{KdevAppx}) with (\ref{Hamiltonian}), we see that the only strategy for obtaining a Hamiltonian in the form (\ref{Hamiltonian}) is to choose $\vec{\mathcal{A}}_{\perp}=0$.  (This has the further advantage of ensuring that the canonical and mechanical momenta coincide.)  It follows that we obtain the Hamiltonian (\ref{Hamiltonian}) by setting $\mathcal{A}_s(X,Y,s)=\tilde{t}U(X_N,Y_N)$, where
\begin{equation}
\tilde{t}=\frac{\tau c^2}{\beta(s)},\quad X_N=\frac{X}{c\sqrt{\beta(s)}},\quad Y_N=\frac{Y}{c\sqrt{\beta(s)}}. \label{scaling}
\end{equation}  
(Note that in (\ref{scaling}), $c$ and $\beta$ denote the parameters appearing in (\ref{Hamiltonian}), not the usual relativistic factors.)
We briefly examine some consequences.  First, note that setting $\vec{\mathcal{A}}_{\perp}=0$ gives $B_x=\partial_yA_s$, $B_y=-\partial_xA_s$, and $B_s=0$ everywhere.
This implies that $\nabla\cdot\vec{B}=0$ as expected, but note that
\begin{align}
\nabla\times\vec{B}&=-\hat{x}(\partial_sB_y)+\hat{y}(\partial_sB_x)+\hat{s}(\partial_xB_y-\partial_yB_x) \notag \\
&=\hat{x}(\partial_s\partial_xA_s)+\hat{y}(\partial_s\partial_yA_s)-\hat{s}(\partial_x^2A_s+\partial_y^2A_s)=\nabla_{\perp}(\partial_sA_s), \label{Bcurl}
\end{align}
where in the last step we used the fact that $(\partial_x^2+\partial_y^2)A_s=0$ by (\ref{Uproperties}).  For (\ref{Bcurl}) to vanish requires that $\nabla_{\perp}(\partial_sA_s)=0$, which is equivalent to the condition that $\partial_sB_x=\partial_sB_y=0$ (the transverse magnetic field must be independent of $s$).  However, the $s$-dependence of $A_s$ is determined by the betatron amplitude $\beta(s)$ through (\ref{Hamiltonian}), so this condition is not met in general.  That is, there is no valid magnetic vector potential producing the Hamiltonian (\ref{Hamiltonian}) {\it except} in the approximate 2D limit that $\partial_sA_s\rightarrow 0$, or equivalently $\beta'(s)\rightarrow 0$.  In the following section, we consider this limiting case in detail.

\section{Solution for the 2D Potentials and Magnetic Field}
In this section, we describe how to determine the vector potential and the associated magnetic field for the nonlinear insert in the idealized 2D limit.
In addition, we would also like to determine a 2D magnetic scalar potential $\psi$ such that $\vec{B}_{\perp}=-\nabla_{\perp}\psi$.  (See \cite{Halbach}.)  For convenience, we will normalize $\psi$ by $B\rho$, so that $\psi$ is dimensionless.  Likewise, we use the dimensionless transverse coordinates $X_N$ and $Y_N$ of (\ref{scaling}) throughout, but we will {\it denote these by the lowercase variables $x$ and $y$ to simplify notation}.  Note that Maxwell's equations imply that both $\mathcal{A}_s$ and $\psi$ must be harmonic, that is:  $(\partial_{x}^2+\partial_y^2)\mathcal{A}_s=0$ and $(\partial_{x}^2+\partial_y^2)\psi=0$.  In addition, the vector and scalar potentials must be related through:  $\partial_{{y}}\mathcal{A}_s=-\partial_{{x}}\psi$ and $\partial_{{x}}\mathcal{A}_s=\partial_{{y}}\psi$.  

If we set ${x}+i{y}=z\in\mathbb{C}$, these are exactly the Cauchy-Riemann equations for the real and imaginary parts of a single analytic function $F=\mathcal{A}_s+i\psi$.  That is, $\psi$ must be a harmonic conjugate of $\mathcal{A}_s$.
To ensure that $\psi$ exists, let $D$ be any open, simply connected subset $D\subset\mathbb{C}$ on which $\mathcal{A}_s$ is harmonic; then Theorem 1 of the Appendix states that there exists a harmonic conjugate of $\mathcal{A}_s$ on $D$ that is unique up to a constant.  The optimal choice of $D$ will depend on the location of any singularities in the potential, and this will be described in the following section.

\subsection{Potentials in the complex plane}
We search for a vector potential $\mathcal{A}_s$ defined on an open, simply connected domain $D$ in the plane such that:  1) $\mathcal{A}_s$ is harmonic on $D$, and 2) $\mathcal{A}_s$ satisfies the Bertrand-Darboux equation on $D$:
\begin{equation}
(\partial_x^2+\partial_y^2)\mathcal{A}_s=0,\quad\quad xy(\partial_x^2-\partial_y^2)\mathcal{A}_s+(y^2-x^2+1)\partial_x\partial_y\mathcal{A}_s+3y\partial_x\mathcal{A}_s-3x\partial_y\mathcal{A}_s=0. \label{BDeq}
\end{equation}
We claim that these two conditions are equivalent to the following complex ODE for a single analytic function $F=\mathcal{A}_s+i\psi$:
\begin{equation}
(z^2-1)\frac{d^2F}{dz^2}+3z\frac{dF}{dz}+2\tilde{t}=0, \quad\quad (z\in D)\label{complexeq}
\end{equation}
where $\tilde{t}\in\mathbb{R}$ is an arbitrary constant.
To see that (\ref{complexeq}) implies (\ref{BDeq}), suppose that $F$ is an analytic solution of (\ref{complexeq}) on $D$.  The complex derivative is given in terms of $x$ and $y$ as:
\begin{equation}
\frac{dF}{dz}=\frac{1}{2}(\partial_x-i\partial_y)F.
\end{equation}
Using the Cauchy-Riemann equations, it follows that the action of the complex derivatives $d/dz$ and $d^2/dz^2$ on $F=\mathcal{A}_s+i\psi$ can be expressed in terms of the real part of $F$ as:
\begin{equation}
\frac{dF}{dz}=(\partial_x-i\partial_y)\mathcal{A}_s,\quad\quad \frac{d^2F}{dz^2}=\frac{{1}}{2}(\partial_x^2-\partial_y^2-2i\partial_x\partial_y)\mathcal{A}_s. \label{complexderivs}
\end{equation}
Substituting these expressions into (\ref{complexeq}) and taking the imaginary part gives exactly the rightmost equation in (\ref{BDeq}).  Likewise, since $\mathcal{A}_s=\mathcal{R}e (F)$ for $F$ analytic on $D$, it follows that $\mathcal{A}_s$ is harmonic on $D$, and so both equations in (\ref{BDeq}) are satisfied.  Conversely, if $\mathcal{A}_s$ satisfies (\ref{BDeq}), then $\mathcal{A}_s$ is the real part of some analytic function $F=\mathcal{A}_s+i\psi$ on $D$.  For this function $F$, it follows from (\ref{complexderivs}) that the rightmost equation in (\ref{BDeq}) is exactly the imaginary part of (\ref{complexeq}).  Thus, the imaginary part of (\ref{complexeq}) must vanish.  Since the left-hand side of (\ref{complexeq}) is analytic on $D$, the vanishing of the imaginary part of (\ref{complexeq}) implies that the real part must be equal to a constant (independent of $z$).  This shows that (\ref{complexeq}) and (\ref{BDeq}) are equivalent.  Thus, determination of the desired quantities $\mathcal{A}_s$ and $\psi$ is reduced to solution of the complex ordinary differential equation (\ref{complexeq}).

Next, set $G=F'$, and note that (\ref{complexeq}) can be written as the first-order linear inhomogeneous system:
\begin{equation}
\frac{dG}{dz}=\frac{3z}{1-z^2}G+\frac{2\tilde{t}}{1-z^2},\quad\quad \frac{dF}{dz}=G. \label{system}
\end{equation}
Note that the right-hand side of the leftmost equation has poles at $z=\pm 1$, which we would like to avoid.  Set $E=\{z\in\mathbb{C}:\mathcal{I}m(z)=0,\quad |\mathcal{R}e(z)|\geq 1\}$, and choose the domain $D$ such that $D=\mathbb{C}\setminus E$.  (See Fig. \ref{Domain}.)  Then $D$ is open and simply connected, so Theorem 2 of the Appendix implies that an analytic solution exists everywhere on $D$.  Furthermore, this solution will be unique if we specify the values of $F$ and $G=F'$ at the origin.

One may solve (\ref{system}) directly as follows.  Use an integrating factor $\mu(z)=(1-z^2)^{3/2}$ to give for $z\in D$:
\begin{equation}
\frac{d}{dz}(\mu G)=\mu G'+\mu'G=(1-z^2)^{3/2}\left(G'+\frac{3z}{z^2-1}G\right)=(1-z^2)^{3/2}\left(\frac{2\tilde{t}}{1-z^2}\right)=2\tilde{t}(1-z^2)^{1/2}. \label{ifactor}
\end{equation}
It is natural to set $G(0)=F'(0)=0$.  (We will see that this is equivalent to requiring that the on-axis magnetic field vanishes.)  Integrating (\ref{ifactor}) along any path in $D$ from 0 to $z\in D$ gives:
\begin{equation}
\mu(z)G(z)=2\tilde{t}\int_0^z(1-\zeta^2)^{1/2}d\zeta=\tilde{t}\left(z\sqrt{1-z^2}+\operatorname{arcsin}(z)\right). \label{integral1}
\end{equation}
Next, we also assume that $F(0)=0$ and integrate $G=F'$ along any path in $D$ from 0 to $z\in D$ to give:
\begin{equation}
F(z)=\tilde{t}\int_0^z\left(\frac{\zeta}{1-\zeta^2}+\frac{\operatorname{arcsin}(\zeta)}{(1-\zeta^2)^{3/2}}\right)d\zeta=\left(\frac{\tilde{t}z}{\sqrt{1-z^2}}\right)\operatorname{arcsin}(z). \label{integral2}
\end{equation}
Thus, the unique solution of (\ref{complexeq}) satisfying $F(0)=0$ and $F'(0)=0$ is given by the simple expression:
\begin{equation}
F(z)=\left(\frac{\tilde{t}z}{\sqrt{1-z^2}}\right)\operatorname{arcsin}(z)\quad\quad (z\in D). \label{ComplexResult}
\end{equation}
We use the principal branch for all functions, so branch cuts occur along $E$, which is outside the domain $D$ of interest (Fig. \ref{Domain}).  The paths occurring in the integrals (\ref{integral1}-\ref{integral2}) could be taken as straight line segments.  Note that (\ref{ComplexResult}) is analytic on $D$, as expected, so both $\mathcal{A}_s=\mathcal{R}e(F)$ and $\psi=\mathcal{I}m(F)$ are harmonic on $D$, as required. 
\begin{figure}
\begin{center}
\resizebox{3in}{!}{\includegraphics{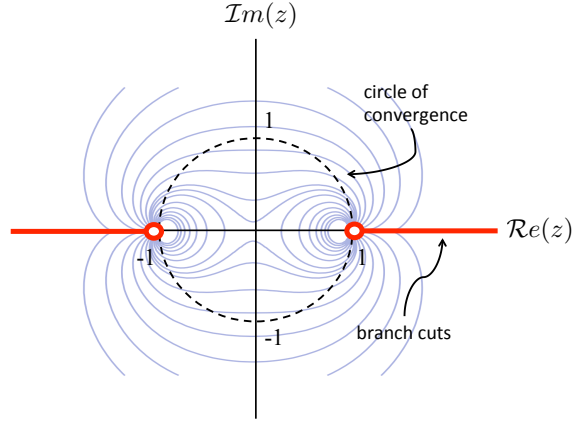}}
\caption{\label{Domain}  The domain of definition $D$ for the complex potential (\ref{ComplexResult}).   (Red lines are excluded.)  Singularities occur at $z=\pm 1$, and $\psi=\mathcal{I}m(F)$ is discontinuous across the two branch cuts.  The multipole series converges within the dashed circle.  The blue curves denote the corresponding magnetic field lines.}
\end{center}
\end{figure}

The power series for $F=\mathcal{A}_s+i\psi$ about the origin takes the form:
\begin{equation}
F(z)=\tilde{t}\sum_{n=1}^{\infty}\frac{2^{2n-1}n!(n-1)!}{(2n)!}z^{2n},\quad |z|<1. \label{Fseries}
\end{equation}
This can be verified to all orders without explicitly computing the series from (\ref{ComplexResult}).  To see this, note that (\ref{Fseries}) immediately implies that $F(0)=0$ and $F'(0)=0$.  Substituting (\ref{Fseries}) into (\ref{complexeq}) gives a sequence of recursion relations for the series coefficients, which are correctly satisfied to all orders.  By the uniqueness provided by Theorem 2 of the Appendix, (\ref{Fseries}) must therefore give the power series representation for (\ref{ComplexResult}), whose radius of convergence is set by the singularities at $z=\pm 1$.  Direct verification shows that this is identical to the power series given in Eqs. (1-2) of \cite{OShea}.

Direct verification also shows that taking the real part of (\ref{ComplexResult}) reproduces the potential defined in Sec.V.A. of \cite{Danilov}, provided we consider the nominal case that $\tilde{b}\equiv b/t=-\pi/2$ and $\tilde{d}\equiv d/t=0$ (no magnetic field at the origin).
If we instead allow a nonvanishing magnetic field at the origin, so that $F'(0)\in\mathbb{C}$ is nonzero, following an identical procedure yields a slightly modified solution of (\ref{complexeq}):
\begin{equation}
F(z)=\left(\frac{z}{\sqrt{1-z^2}}\right)\left(\tilde{t}\operatorname{arcsin}(z)+F'(0)\right)\quad\quad (z\in D).
\end{equation}
The real and imaginary parts of $F'(0)$ then determine the two real constants $b$ and $d$ appearing in Eq. (21) of \cite{Danilov}, with $F'(0)=0$ if and only if $b/t=-\pi/2$ and $d/t=0$.

\subsection{Magnetic field and thin-lens kick}
Recall that the magnetic field components are given by $B_x=\partial_yA_s$ and $B_y=-\partial_xA_s$, so
\begin{equation}
B_x+iB_y=\left(\frac{\partial}{\partial y}-i\frac{\partial}{\partial x}\right)A_s=\left(\frac{B\rho}{c\sqrt{\beta}}\right)\left(\frac{\partial}{\partial Y_N}-i\frac{\partial}{\partial X_N}\right)\mathcal{A}_s
=-i\left(\frac{B\rho}{c\sqrt{\beta}}\right)\frac{dF}{dz}^*,
\end{equation}
where the star denotes complex conjugation.  It is therefore natural to define a dimensionless magnetic field by:
\begin{equation}
\mathcal{B}=\left(\frac{c\sqrt{\beta}}{B\rho}\right)(B_x+iB_y),
\end{equation}
so that
\begin{equation}
\mathcal{B}^*=i\frac{dF}{dz}=i\tilde{t}\left\{\frac{z}{1-z^2}+\frac{\operatorname{arcsin}(z)}{(1-z^2)^{3/2}}\right\}.
\end{equation}
In particular, the function $\mathcal{B}$ is also analytic on $D$, and its power series about the origin is given for $|z|<1$ by differentiating the power series (\ref{Fseries}).

Finally, let us determine the change in transverse momentum that an on-energy particle experiences when crossing a thin nonlinear segment of length $\Delta s$.  From (\ref{KdevAppx}) we see that (since $\vec{\mathcal{A}}_{\perp}=0$):
\begin{equation}
\frac{dP_x}{ds}=-\frac{\partial H_{\perp}}{\partial x}=\frac{1}{c\sqrt{\beta}}\frac{\partial\mathcal{A}_s}{\partial X_N},\quad\quad \frac{dP_y}{ds}=-\frac{\partial H_{\perp}}{\partial y}=\frac{1}{c\sqrt{\beta}}\frac{\partial\mathcal{A}_s}{\partial Y_N},
\end{equation}
so the change in transverse momentum over a thin segment of length $\Delta s$ is given by:
\begin{equation}
\Delta P_x+i\Delta P_y=\left(\frac{\Delta s}{c\sqrt{\beta}}\right)\left(\frac{\partial}{\partial X_N}+i\frac{\partial}{\partial Y_N}\right)\mathcal{A}_s=\left(\frac{\Delta s}{c\sqrt{\beta}}\right)\frac{dF}{dz}^*. \label{momentum}
\end{equation}
It is therefore natural to define a dimensionless momentum kick $p$ and a dimensionless segment length $\sigma$ by:
\begin{equation}
p=\Delta P_x+i\Delta P_y,\quad\quad \sigma=\frac{\Delta s}{c\sqrt{\beta}}.
\end{equation}
Then (\ref{momentum}) gives:
\begin{equation}
p^*=\sigma\frac{dF}{dz}.
\end{equation}
In particular, this justifies the nominal choice of setting $F'(0)=0$; this choice ensures that an on-axis particle experiences no transverse deflection.

\section{Solution for the Two Invariants of Motion}
In this section, we find the functions $U$ and $W$ appearing in (\ref{Htransform}-\ref{Itransform}) that are needed to describe the two invariants of motion.  Continuing the precedent set in Section 4, we use the dimensionless coordinates defined by (\ref{scaling}) throughout.  Recall from the discussion leading to (\ref{scaling}) that the potential $U$ is given in terms of the magnetic vector potential $\mathcal{A}_s$ by $\mathcal{A}_s=\tilde{t}U$, so it follows from (\ref{ComplexResult}) that $U$ is given explicitly by:
\begin{equation}
U(x,y)=\mathcal{R}e\left(\frac{z}{\sqrt{1-z^2}}\operatorname{arcsin}(z)\right),\quad z=x+iy. \label{Uresult}
\end{equation}
To determine the function $W$, we must solve the system given in (\ref{Weq}):
\begin{align}
W_x=2(y^2+1)U_x-2xyU_y,\quad\quad W_y=-2xy U_x+2x^2 U_y. \label{Weq2}
\end{align}
Consider a vector field on the domain $D$ (Fig. \ref{Domain}) whose components are given by the two right-hand sides of (\ref{Weq2}).  Since $U$ is harmonic on $D$, the two right-hand sides of (\ref{Weq2}) are smooth on $D$.  Furthermore, the fact that $U$ satisfies the Bertrand-Darboux equation implies that this vector field is conservative.  Since $D$ is simply connected, it follows that there exists a smooth function $W$ on $D$ satisfying (\ref{Weq2}), and this function is unique up to a constant.  In particular, the value of $W$ at any point in $D$ can be obtained by taking the path integral of $(W_x,W_y)$ from the origin to that point, where we assume $W(0,0)=0$.
The key observation is to note from (\ref{Weq2}) that:
\begin{equation}
xW_x+yW_y=2xU_x. \label{Dplane}
\end{equation}
We perform the path integral from $(0,0)$ to $(x_d,y_d)$ in polar coordinates along a straight segment $C$, so that the angle $\phi$ is fixed along $C$ and:
\begin{equation}
W(x_d,y_d)=\int_C (W_xdx+W_ydy)=\int_0^{\rho_d} (W_x\cos\phi+W_y\sin\phi)d\rho=\int_0^{\rho_d}2U_x\cos\phi d\rho,
\end{equation}
where in the last step we used (\ref{Dplane}).  Then, noting that $\tilde{t}U_x=\mathcal{R}e(F')$ we have:
\begin{equation}
\tilde{t}W(x_d,y_d)=2\cos\phi\cdot\mathcal{R}e\int_0^{\rho_d}F'(\rho e^{i\phi})d\rho=2\cos\phi\cdot\mathcal{R}e \left(e^{-i\phi}\int_0^{z_d} F'(z)dz\right),
\end{equation}
which is now given as a path integral of $F'$ in the complex plane.  Thus, since $F(0)=0$ it follows that:
\begin{equation}
\tilde{t}W(x,y)=2\cos\phi\cdot\mathcal{R}e[e^{-i\phi}F(z)],\quad z=x+iy.
\end{equation}
Eliminating the dependence on $\phi=\arg(z)$ gives the following equivalent form:
\begin{align}
\tilde{t}W(x,y)=\mathcal{R}e [(1+z^*/z)F(z)],\quad z=x+iy. \label{WResult1}
\end{align}
Finally, substituting our solution (\ref{ComplexResult}) for the complex potential $F$ into (\ref{WResult1}) gives the desired result:
\begin{equation}
W=\mathcal{R}e\left(\frac{z+z^*}{\sqrt{1-z^2}}\arcsin(z)\right). \label{WResult2}
\end{equation}
Note that the complex function appearing within parentheses in (\ref{WResult2}) is not analytic, due to its dependence on $z^*$.  However, it is not difficult to verify that $W$ itself is {\it real-analytic} on $D$ when considered as a function of the two real variables $x$ and $y$.

Direct verification shows that the two dimensionless invariants $H_N$ and $I_N$ given in (\ref{Htransform}-\ref{Itransform}), in which $U$ and $W$ are described by (\ref{Uresult}) and (\ref{WResult2}), correspond to the quantities $H_N=H/c^2$ and $I_N=I/c^4$, where $H$ and $I$ are the invariants given in Section V.A. of \cite{Danilov}.

\section{Summary of Results}
Below we summarize the final results in terms of physically relevant parameters.  Magnetic vector and scalar potentials are defined so that $\vec{B}=\nabla\times\vec{A}=-\nabla\psi$.

Summary of dimensionless quantities:
\begin{align}
& F=\frac{A_s+i\psi}{B\rho},\quad z=\frac{x+iy}{c\sqrt{\beta}},\quad p=\frac{\Delta p_x+i\Delta p_y}{p^0}, \label{dimensionlessquantities} \\
& \mathcal{B}=\left(\frac{c\sqrt{\beta}}{B\rho}\right)(B_x+iB_y),\quad \sigma=\frac{\Delta s}{c\sqrt{\beta}},\quad \tilde{t}=\frac{\tau c^2}{\beta}. \notag
\end{align}
Summary of potentials, fields, and momentum kick through a segment:
\begin{equation}
F(z)=\left(\frac{\tilde{t}z}{\sqrt{1-z^2}}\right)\operatorname{arcsin}(z),\quad \mathcal{B}^*=i\frac{dF}{dz},\quad p^*=\sigma\frac{dF}{dz}. \label{summarypotentials}
\end{equation}
Summary of the invariants of motion (normalized coordinates are defined in (\ref{normalizing})):
\begin{align}
H_N&=\frac{1}{2}(P_{xN}^2+P_{yN}^2+X_N^2+Y_N^2)-\tau U(X_N,Y_N), \quad\quad U=\mathcal{R}e\left(\frac{z}{\sqrt{1-z^2}}\operatorname{arcsin}(z)\right),\\ 
I_N&=(X_NP_{yN}-Y_NP_{xN})^2+P_{xN}^2+X_N^2-\tau W(X_N,Y_N), \quad W=\mathcal{R}e\left(\frac{z+z^*}{\sqrt{1-z^2}}\operatorname{arcsin}(z)\right).  \notag
\end{align}
A summary of physical parameters is provided in Table \ref{table:parameters}.
\begin{table}[htbp]
\renewcommand{\arraystretch}{1.3}
\caption{Physical parameters used in the definitions (\ref{dimensionlessquantities}-\ref{summarypotentials}).}
\label{table:parameters}
\centering
\begin{tabular}{llcl}
\hline
Symbol & Description & Unit \\
\hline
$p^0$ & design momentum & eV/c \\
$B\rho$ & magnetic rigidity & T-m \\
$\tau$ & dimensionless strength of NLI & n/a \\
$c$ & scale parameter for NLI & m$^{1/2}$ \\
$B_x$, $B_y$ & magnetic field of NLI & T \\
$A_s$, $\psi$ & vector, scalar potential of NLI & T-m \\
$\beta$ & local betatron amplitude & m \\
$\Delta s$ & length of segment of NLI & m \\
\hline
\end{tabular}\\
\end{table}

\section{Conclusions}
We have shown that the 2D magnetic field, vector potential, scalar potential, and transverse momentum kick across a segment of the IOTA nonlinear insert can be obtained in the concise form (\ref{summarypotentials}) using methods independent of Sec. V.A. of \cite{Danilov}.  This representation provides insight into the analytical structure of the magnetic potential(s), and it can be used to implement efficient numerical tracking through the nonlinear insert in any language that supports complex arithmetic.  In particular, each complex function in (\ref{summarypotentials}) is well-behaved except near the location of the two branch cuts (Fig. \ref{Domain}), which lie well outside the region of interest for particle tracking.  The multipole series is easily recovered.  A similar representation has also been obtained for the second invariant of motion, so that both 1) particle tracking and 2) evaluation of the two invariants can now be performed without using the transformation (\ref{transformation}), which has previously been found to introduce numerical difficulties for particles crossing the midplane.

\section{Acknowledgments}
This work was was supported by the Director, Office of Science, Office of High Energy Physics, of the U.S. Department of Energy under Contract No. DE-AC02-05CH11231.

\section{Appendix}
The following two theorems play a role in the text, and proofs can be found in standard references such as \cite{Conway} and \cite{Coddington}, respectively.  The topology of the domain is important; without the hypothesis that the domain $D$ is simply connected, both theorems are false.

{\it Theorem 1}: // Let $D$ be an open, simply connected subset of $\mathbb{C}$.  If $u:D\rightarrow\mathbb{R}$ is harmonic, then there exists a harmonic function $v:D\rightarrow\mathbb{R}$ such that $f=u+iv$ is analytic on $D$.  The function $v$ is said to be a {\it harmonic conjugate} of $u$.  Any two harmonic conjugates of $u$ differ by at most a complex constant.

{\it Theorem 2}: // Let $D$ be an open, simply connected subset of $\mathbb{C}$.  Suppose $A:D\rightarrow\mathbb{C}^{n\times n}$ is an $n$-by-$n$ matrix function and $b:D\rightarrow\mathbb{C}^n$ is an $n$-dimensional vector function, each of whose components are analytic on $D$, and let $z_0\in D$.  Then the inhomogeneous linear system
\begin{equation}
w'=A(z)w+b(z)
\end{equation}
has a unique analytic solution $w:D\rightarrow\mathbb{C}^n$ that satisfies the initial value condition
\begin{equation}
w(z_0)=w_0,
\end{equation}
where $w_0\in\mathbb{C}^n$.

\end{document}